\documentclass[graybox]{svmult}

\usepackage{mathptmx}       
\usepackage{helvet}         
\usepackage{courier}        
\usepackage{type1cm}        
\usepackage{makeidx}         
\usepackage{graphicx}        
\usepackage{multicol}        
\usepackage[bottom]{footmisc}

\makeindex             

\begin{document}

\title*{Dark halos as seen with gravitational lensing}
\author{Konrad Kuijken}
\institute{Leiden Observatory, Leiden University, PO Box 9513, 2300CA Leiden, The Netherlands, \email{kuijken@strw.leidenuniv.nl}
}
%
%
\maketitle

\abstract{
Dark matter is an important ingredient of galaxies, as was recognised
early on by Ken Freeman himself! Evidence for dark
matter halos is still indirect, based on analysing motions of tracers such
as gas and stars. In a sense the visible galaxy is the mask through
which we can study the dark matter.
Light rays are also sensitive to gravitational fields, and dark haloes
cause observable gravitational lensing effects. There are three
regimes: microlensing (which probes the clumpiness of dark matter
haloes), strong lensing (sensitive to the inner mass distribution) and
weak lensing (which can probe haloes out to 100s of kpc from the
center). This review will concentrate on weak lensing, and describe a
new survey, the Kilo-Degree Survey (KiDS) that is designed to study
galaxy halo masses, extents and shapes as a function of environment,
galaxy type and redshift.}

\section{Ken Freeman's Dark Side}
\label{sec:1}
Dark matter haloes are an important ingredient of galaxies, and indeed
of the universe as a whole. They dominate the mass of material in
galaxies, and drive the gravitational instability that leads to galaxies
condensing and clustering in the first place.

The most convincing indications for dark matter haloes in galaxies are
still from rotation curves of extended HI disks. In fact, the first
mention of extra mass in the literature appears to be in
\cite{freeman1970}, where exponential galaxy disks are described,
their gravitational potentials are calculated analytically, and where,
(in an appendix!), the calculations are compared with then available HI
rotation curves of four bulgeless galaxies. For NGC 300, Ken
prophetically wrote,
without further comment, {\em ``If the HI rotation curve is correct, then
there must be undetected matter beyond the optical extent of NGC 300;
its mass must be at least of the same order as the mass of the
detected galaxy''}.

Further influential work on dark matter haloes includes the work with
van der Kruit (described elsewhere in this volume) on understanding
the internal dynamics of stellar disks and hence their contribution to
the total mass budget of the galaxy (are disks' contributions to the
rotation curve ``maximal'' or not?), which included some of the first
deep spectroscopy and integrated-light velocity dispersion
measurements of galaxy disks ever attempted. See van der Kruit's
chapter for more details.

A final contribution to be highlighted here is Ken's part in the MACHO
project \cite{macho1993}, which used the (now sadly burnt down) Mt. Stromlo 50-inch
telescope to monitor millions of stars in the Large Magellanic Cloud
for many years and discovered tens of gravitational microlensing
events. This experiment probes the clumpiness of the matter
distribution along the line of sight to the LMC, and the low number of
events seen by MACHO (and by the similar EROS experiment \cite{eros1998}) proved that
the Galactic dark halo does not consist predominantly of compact dark
objects. Any substantial population of dark objects more massive than
Earth is ruled out. 

\section{Gravitational lensing and dark matter}

Even though dynamical tracers were the first to provide evidence for
dark matter, photons are a very useful alternative. Because of their
speed they are only deflected slightly as they pass a massive galaxy
`lens', but the effect is large enough to be measurable and simple
enough to be free from some of the uncertainties inherent in dynamical
tracer studies. A brief comparison of dynamics and lensing as
gravitational potential tracers is given below.
\bigskip

\begin{tabular}{p{0.45\textwidth}@{\hspace{20pt}}p{0.45\textwidth}}
\bf Dynamics&\bf Lensing\\\\
Sensitive to the 3-D potential of the lens & Sensitive
to the 2-D, projected potential on the sky\\\\
Requires assumption of equilibrium & Sensitive to
the instantaneous mass distribution\\\\
Orbit structure unknown & Mass-sheet degeneracy\\\\
Strong lensing: few data points & Good S/N in baryon-dominated
parts\\\\
Weak lensing at large radii requires stacking & Few good tracers at
large radii\\
\end{tabular}

\bigskip
Both techniques have their fundamental limitations. For dynamics this
is the need to understand the orbit structure of the tracer: the most
famous example is the radial orbit anisotropy degeneracy: the same
dispersion profile in a spherical galaxy can be explained with radial
orbit anisotropy, or with a rising mass-to-light ratio \cite{binneymamon1982}. For stellar
tracers disentangling this degeneracy can be complicated requiring
higher-order measurements of the shape of the velocity distribution.
 In the case of lensing, the main uncertainty is the mass-sheet
 degeneracy, which is a transformation of the lens that leaves the observed image
 positions intact but changes the (unobservable) source plane
 scale. The effect on the lens model is to trade some of the lens mass
 for a constant mass sheet, in such a way that the projected mass
 enclosed in the Einstein radius remains unchanged. The mass-sheet degeneracy makes it hard to measure the slope of the projected density of a lens accurately.

Lensing manifests itself in three different settings: microlensing, weak lensing and strong lensing. 
Microlensing is the twinkling that occurs when a source is viewed through a screen of effectively point-mass lenses. Occasionally a lens will pass very close to the line of sight to a source, and temporarily magnify it with a characteristic lightcurve. Observing the statistics and duration of these lightcurves allows the clumpiness of the foreground mass distribution to be studied.
Strong lensing occurs when the lightrays are deflected sufficiently for several images of the same source to be observed. These special geometries allow some of the best mass determinations in cosmology to be made. Finally, weak lensing covers the other cases: a weak deflection that is not sufficient to cause multiple images, but can be detected statistically from the correlated distorting effect it has on the shapes of the sources.

\begin{figure}
\begin{tabular}[b]{c}
\includegraphics[width=0.45\textwidth]{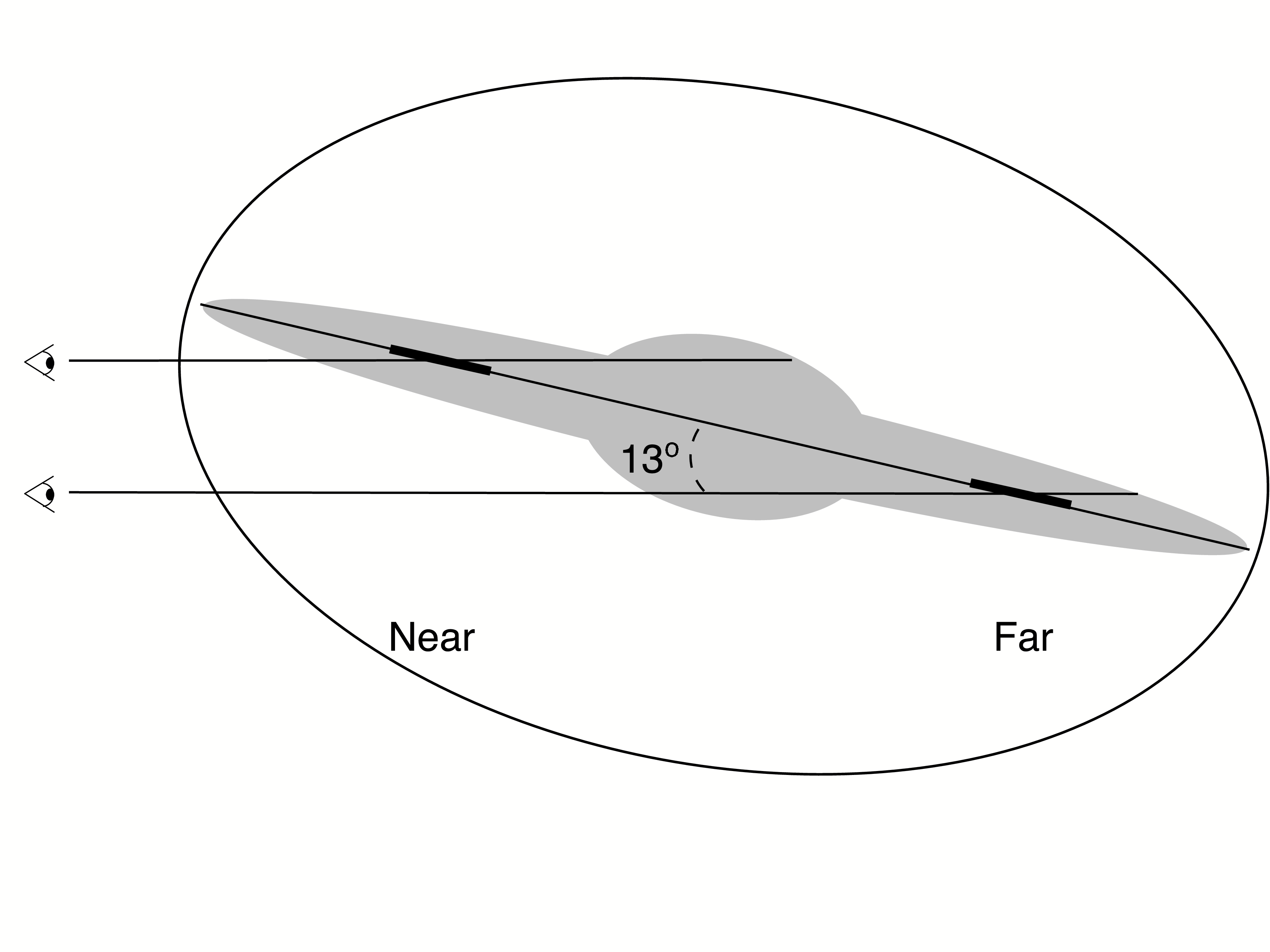}\\
\includegraphics[width=0.45\textwidth]{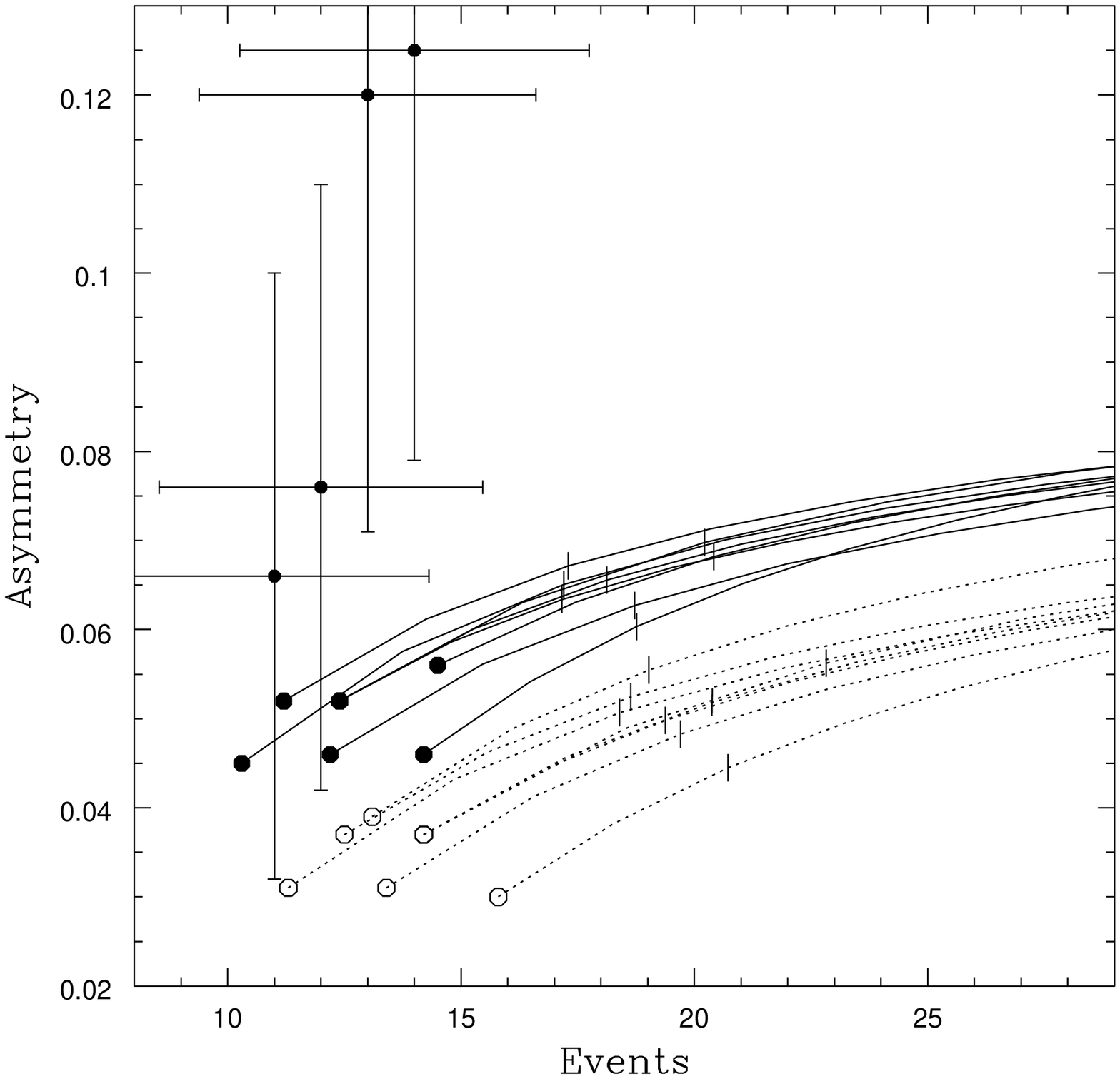}
\end{tabular}
\includegraphics[width=0.5\textwidth]{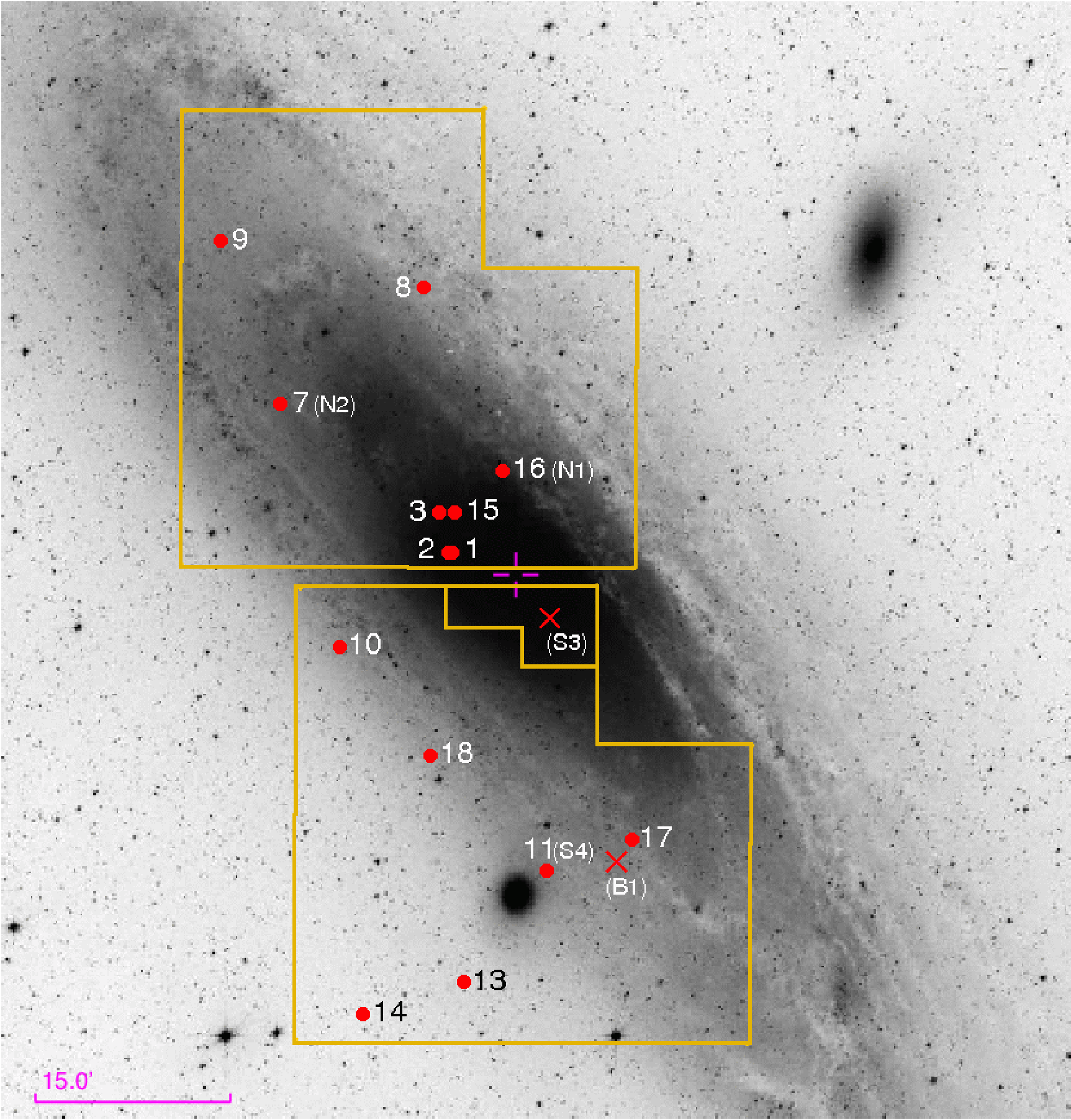}
\caption{Top left: illustration of the asymmetry to be expected in the M31
 halo microlensing rates between the near and far sides of the
 stellar disk. Right: field layout with detected events. Bottom left: Model predictions for number of events, and asymmetry in their distribution (curves) and measurements employing various cuts (data points). Different curves make different assumptions on extinction correction. From \cite{dejongetal2006}.} 
\label{fig:m31}
\end{figure}

\section{Microlensing}

As noted above, the MACHO and EROS experiments established that the
Galactic dark halo is not predominantly composed of massive objects,
at least along the sightline to the Magellanic clouds.

A later study of M31, modestly called the `MEGA' project, found
similar results \cite{dejongetal2006}. In ground-based images of M31
the stars are terribly blended, so that variables can only be found
with a different technique, difference image photometry. This involves
subtracting exposures of the same field after carefully correcting for
seeing and extinction differences, and it has the effect of turning a
crowded field of stars into an uncrowded field of difference flux. The
main difference in the analysis of difference image time series is
that the baseline flux of the star in its unlensed state is lost, but
it turns out that it is still possible to constrain the lens mass
spectrum from such data. In fact, M31 offers several advantages over
LMC studies because the entire M31 halo can be mapped with a single
experiment. In particular, since M31 is inclined to the plane of the
sky, sight lines to the near and far sides of the disk cross different
path lengths of M31's halo. This implies that the two sides of the
disk should see different microlensing rates, making it possible to
distinguish disk self-lensing (which should be the same on both sides)
from halo microlensing. The asymmetric lensing signal also affords (in
principle) a constraint on the shape of the halo.

Results from a 4-year campaign using the Isaac Newton Telescope on La
Palma, using about 200 nights of data, revealed 14 microlensing events
over the face of M31's disk. Analysis is complicated because much of
M31's disk is extincted to varying degree, and because the two fields
monitored did not cover the full galaxy. A full Monte-Carlo simulation
of the experiment concluded that the number of events, and the
asymmetry in their distribution between both sides of the disk, are
fully consistent with the expectation for self-lensing of/by disk and
bulge stars (Fig.~\ref{fig:m31}).

\section{Strong lensing}

Many of the most spectacular astronomical images are caused by strong
gravitational lensing, which splits the light from a distant source
into multiple images. It turns out that these images can provide some
of the best-measured masses in astronomy.

The angle through which a gravitational lens at distance\footnote{All
  distances are defined as angular diameter distances.} $D_l$ moves the image of a
source at distance $D_s$ on the sky is determined by its surface mass density,
scaled by the `critical surface mass density' $\Sigma_{cr}$ given by
\begin{equation}
\Sigma_{cr}={c^2\over 4\pi G} {D_s\over D_l D_{ls}}
\end{equation}
where $D_{ls}$ is the distance of the source from the lens. In the
ideal case in which the source lies directly behind the lens, and the
lens has a circularly symmetric mass distribution on the sky, the
source will be distorted into an `Einstein Ring', inside which the
average surface density is exactly equal to $\Sigma_{cr}$. Even for
not-quite axisymmetric lenses, or not-quite perfect alignment, it is
possible to estimate the Einstein radius of the lens and hence anchor
the mass distribution of the lens.
Typical values for the Einstein radius are an arcsecond, which
corresponds roughly to an effective radius of the lens. In other
words, strong lenses probe the mass distribution in the visible
central parts of massive galaxies. 

The most comprehensive study of the mass distribution of galaxies
from strong lensing has been undertaken by the SLACS team
\cite{slacs3}. They identified candidate strong lenses from the SDSS
spectra, by looking for early-type galaxies with incompatible emission
lines in the spectra that could be due to amplified background
emission-line galaxies. Subsequent HST imaging revealed a very high
success rate of the technique and resulted in a sample of dozens of
new lenses. Unlike the existing lensed-quasar samples, many of these
lenses have resolved sources so that the mass distribution of the lens
can be probed at many positions. A complete modelling effort,
including stellar dynamics measurements of many of the lenses,
resulted in the surprising conclusion that a simple singular
isothermal ellipsoid mass model is able to account for all the data. 

\section{Weak lensing}

Strong lensing probes the central, high surface mass density regions
of galaxies. At larger radii gravitational lensing still occurs, but
not at the strength required to split images. Instead the effect is to
distort the background sources (since different parts of a source are
deflected by different amounts). The leading-order effect of this
distortion is expressed as a two-dimensional symmetric matrix that
describes the mapping from image to (unlensed) source coordinates as a
{\em shear $\gamma$} and a {\em convergence $\kappa$}:
\begin{equation} 
\left(
\begin{array}{c}x_s\cr y_s\end{array}
\right)
=
\left(
\begin{array}{cc} 1-\kappa-\gamma_1 &
  -\gamma_2\cr -\gamma_2&1-\kappa+\gamma_1\end{array}
\right)
\left(
\begin{array}{c}x_i\cr y_i\end{array}
\right)
\equiv
(1-\kappa)
\left(
\begin{array}{cc} 1-g_1 &
  -g_2\cr -g_2&1+g_1\end{array}
\right)
\left(
\begin{array}{c}x_i\cr y_i\end{array}
\right).
\end{equation}
Here $g\equiv\gamma/(1-\kappa)$ is called the {\em reduced shear}, and
it is related to the axis ratio of $b/a$ of the ellipse into which a
circular source is sheared, via $g=(a-b)/(a+b)$. It is the only
quantity that can be measured from galaxy shapes, a fact that is
related to the mass-sheet degeneracy.

The effect of a massive lens can be understood qualitatively as
follows. Gravitational lensing acts to push images on the sky away
from the lens, a consequence of the `detour' that the light has to
make around the lens. This distorts the background sky, effectively
squeezing the sky around a lens radially. The result is a pattern of
tangentially distorted background sources around every foreground mass
concentration.

 Weak lensing consists of extracting the shear field on the sky from
measurements of galaxy shapes, essentially by trying to derive the
shape of the `average' galaxy, which is intrinsically round. This
subject is technically complex, due to the many distorting effects
in astronomical images that have nothing to do with gravitational
lensing, but can be much stronger: smearing by the point spread
function, distortion by the camera optics, pixelation of the image on
the detector, charge transfer inefficiency, etc. The central issue is
the need to average over large numbers of background galaxies in order
to extract the shear against the intrinsic variety of galaxy shapes,
the so-called shape noise. 

Weak lensing is not only an effective way to study the mass
distribution around galaxies and in clusters, it has also been
recognized as a powerful cosmological probe. Lensing strength is
sensitive to the angular-diameter distances of lens and source, and in
combination with redshift measurements can be used to reconstruct the
angular diameter distance-redshift relation, i.e., the expansion
history of the universe. The statistical power of a large, three-D
weak lensing map of the sky is sufficient to yield percent-level
measurements of dark energy equation of state parameters, and is
currently the subject of several design studies for space missions.

\begin{figure}
\includegraphics[width=0.5\textwidth]{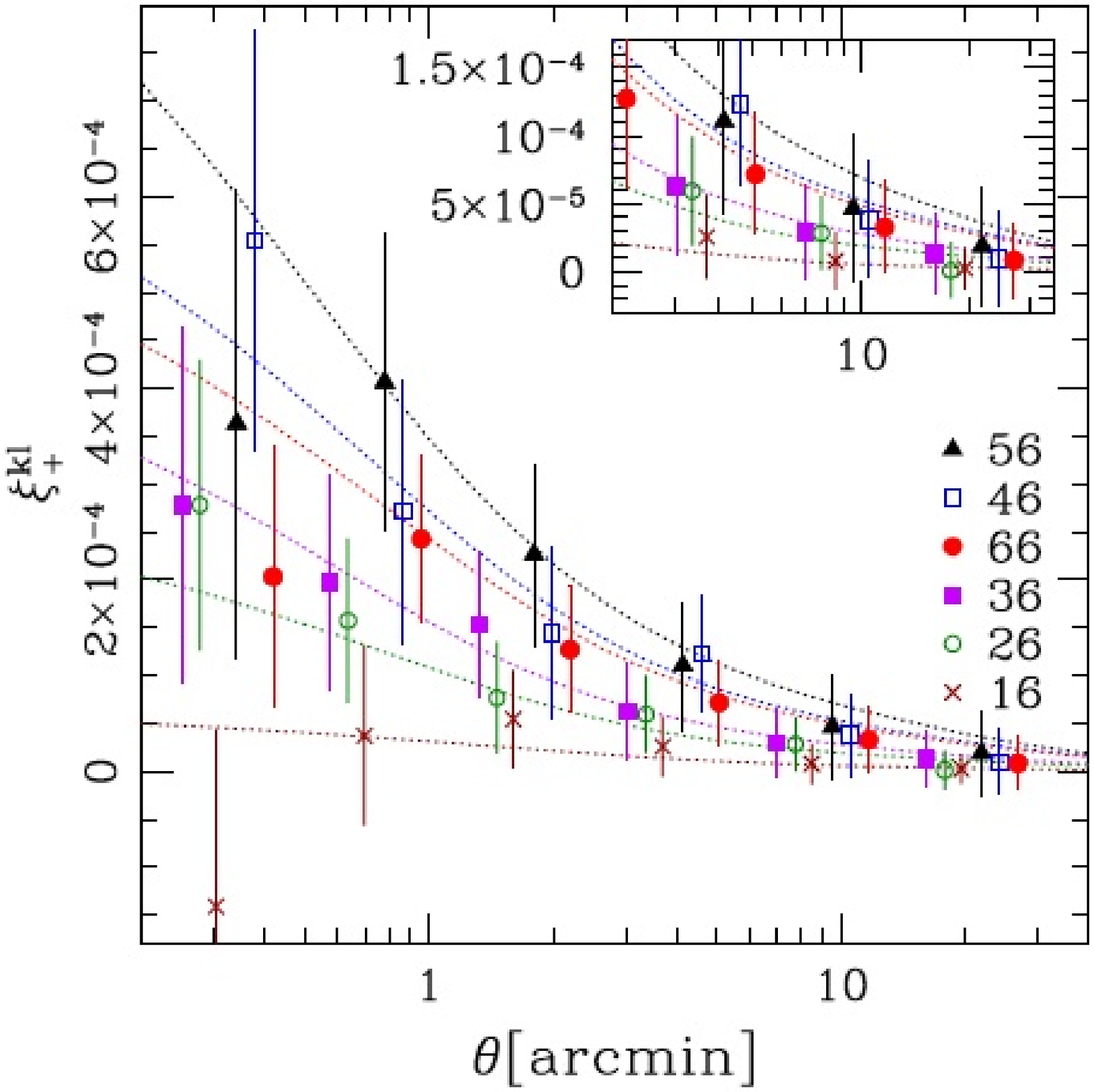}
\includegraphics[width=0.5\textwidth]{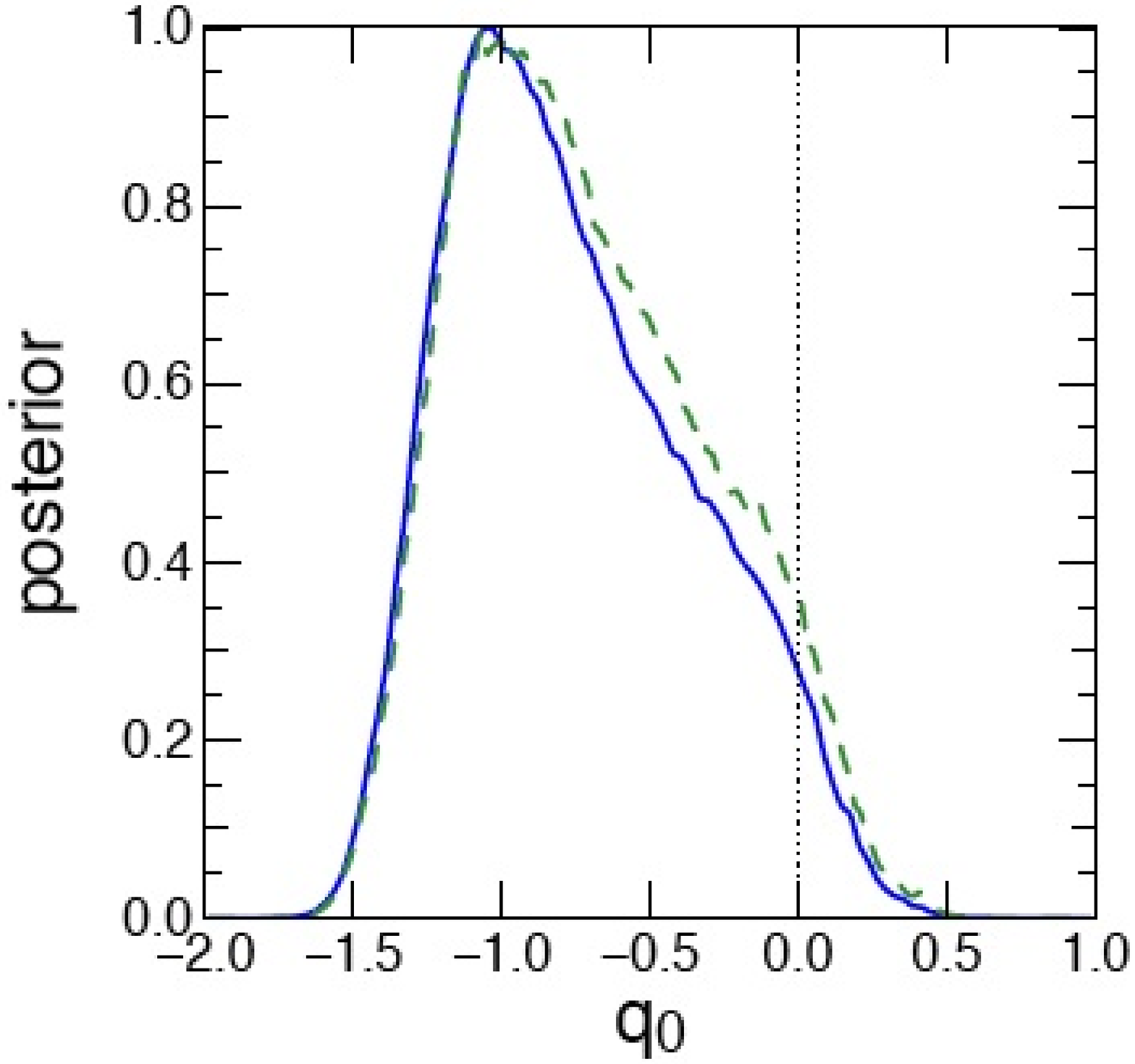}
\caption{Left: cross-correlation functions (symbols) of the shear
 field in the COSMOS field, for different redshift slices. The
 amplitudes of the cross-correlation functions depend only on the
 angular-diameter distances to these redshift slices, and they
 scale as expected (solid lines). Right: Confidence intervals on
 the acceleration parameter, showing that these lensing data strongly
 favour an accelerating universe. From \cite{schrabbacketal2010}.}
\label{fig:cosmos}
\end{figure}
 
Recently we have demonstrated that it is possible to control the
systematic errors to below the statistical errors, for what is
currently the largest space-based weak lensing map, the COSMOS
field \cite{schrabbacketal2010}. Combining photometric redshifts with
weak lensing measurements, we were able to show that the strength of
the gravitational lensing signal increases with source redshift as
expected in the standard cosmological model, and could even add
independent, lensing-based confidence intervals on the acceleration
parameter. Weak lensing is fast becoming established as a reliable
probe for galactic structure and cosmology (Fig.~\ref{fig:cosmos}).

Individual galaxy lenses have too few nearby background galaxies
around them to yield useful mass measurements at large radii: the
shape noise simply overwhelms any shear signal. A clear shear
measurement can only by achieved by stacking large numbers of similar
galaxies. Fortunately, modern wide-field imagers are capable of the
necessary observations.

An impressive demonstration of the power of numbers is provided by the weak lensing analyses of the SDSS image survey. Already the early data release showed a convincing signal \cite{fischer2000}, but the more recent results have refined the result to the point where the average mass distribution can be studied as a function of galaxy stellar mass (derived from SDSS spectroscopy and SED fitting), luminosity, Hubble type and environmental density \cite{mandelbaum2006a}. Tentative results on the shapes of galaxy haloes have also been reported \cite{hoekstra2004, mandelbaum2006b}. A separate study \cite{gavazzi2007}, using HST images, of the SLACS lenses discussed above has also been done, and shows that the singular isothermal lens model continues to fit this population well out to radii of several hundreds kpc.

\subsection{Future directions}

Weak lensing focuses on the most visible effect of lensing, the shear,
but these are not the only measures of lensing strength. Also the 
magnification, and the flexion, are showing promise as independent
measures.

\begin{figure}
\begin{center}\includegraphics[width=0.9\textwidth]{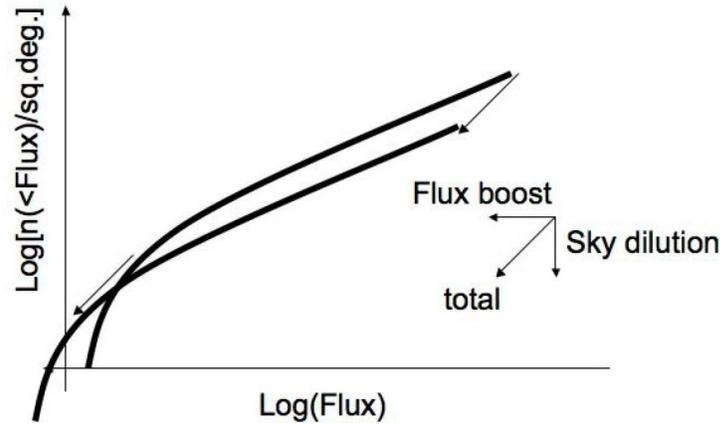}\end{center}
\caption{Illustration of the magnification effect on the luminosity
 function of a population of sources. Depending on the slope of the
 luminosity function the numbers of observed sources per unit area
 may increase or decrease.}
\label{fig:mag1}
\end{figure}

\begin{figure}
\begin{center}
\includegraphics[width=0.49\textwidth]{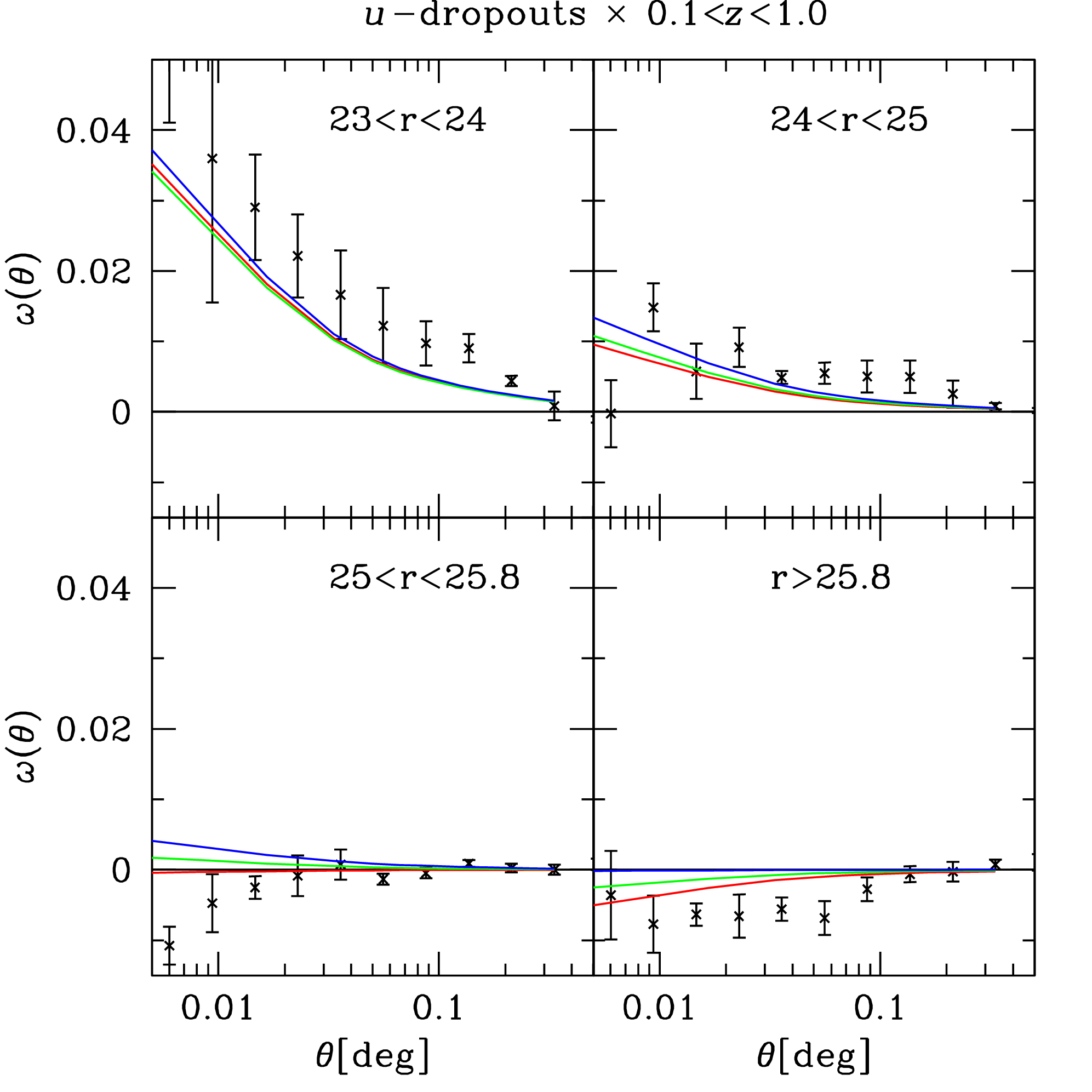}
\includegraphics[width=0.49\textwidth]{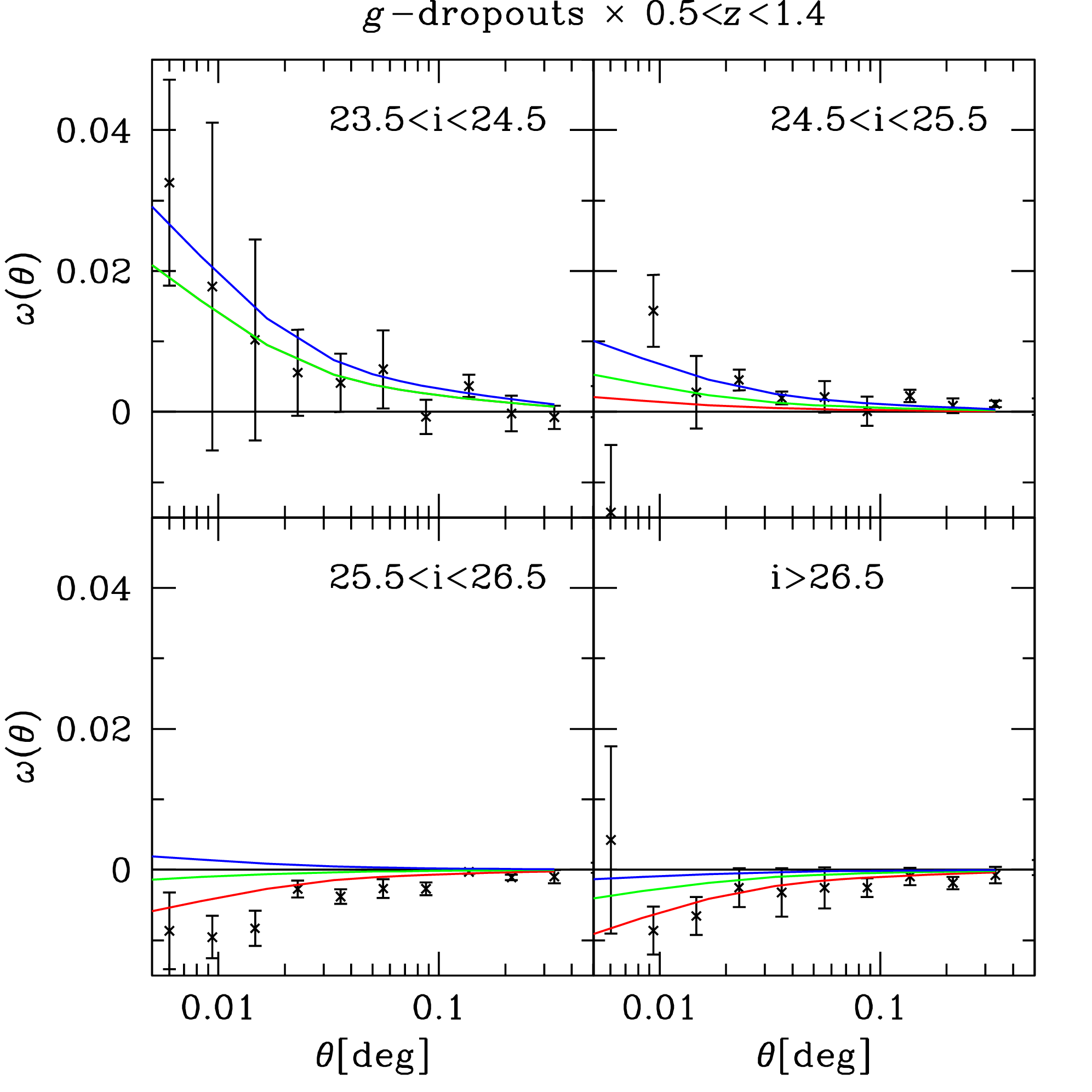}
\end{center}
\begin{center}
\includegraphics[width=0.49\textwidth]{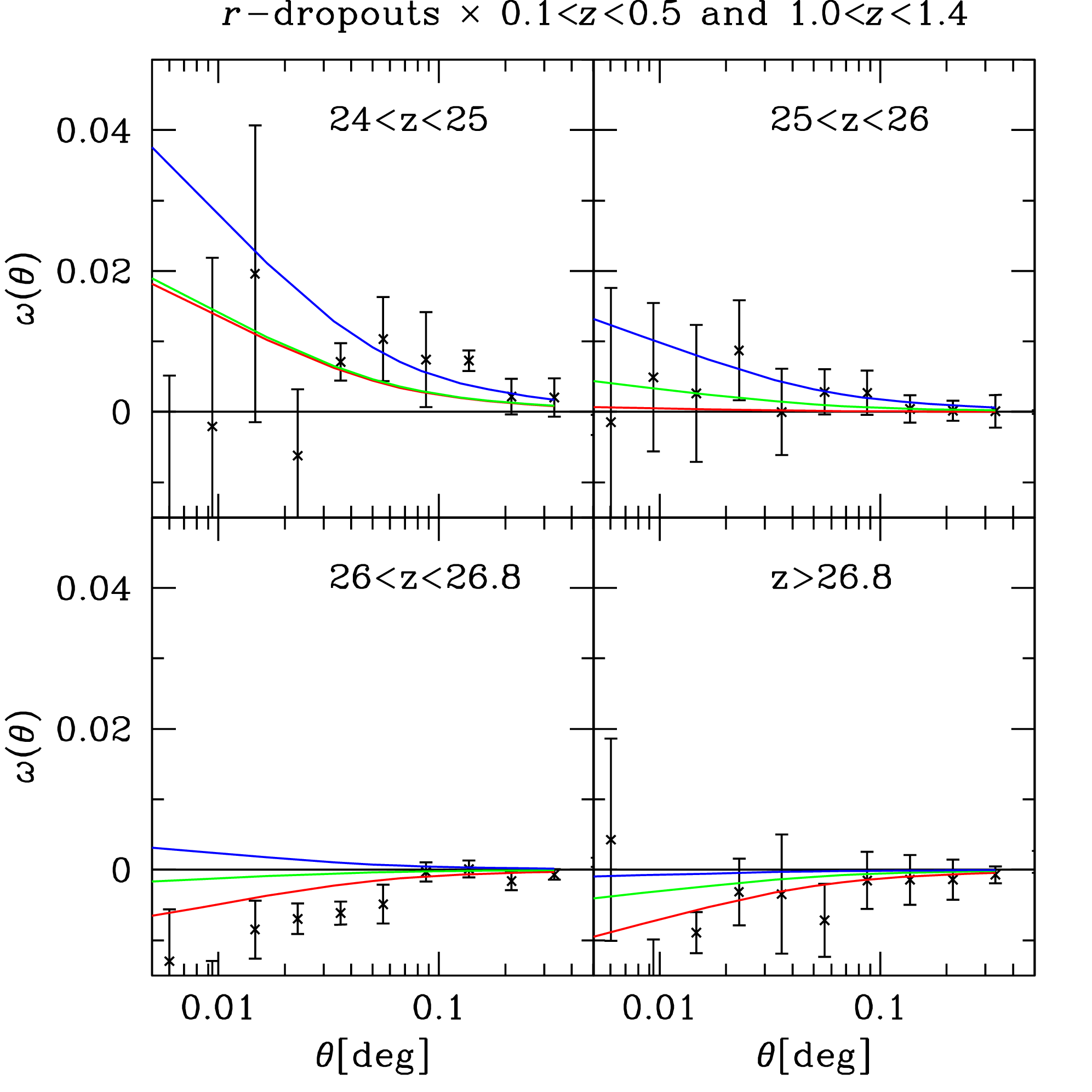}
\includegraphics[width=0.49\textwidth]{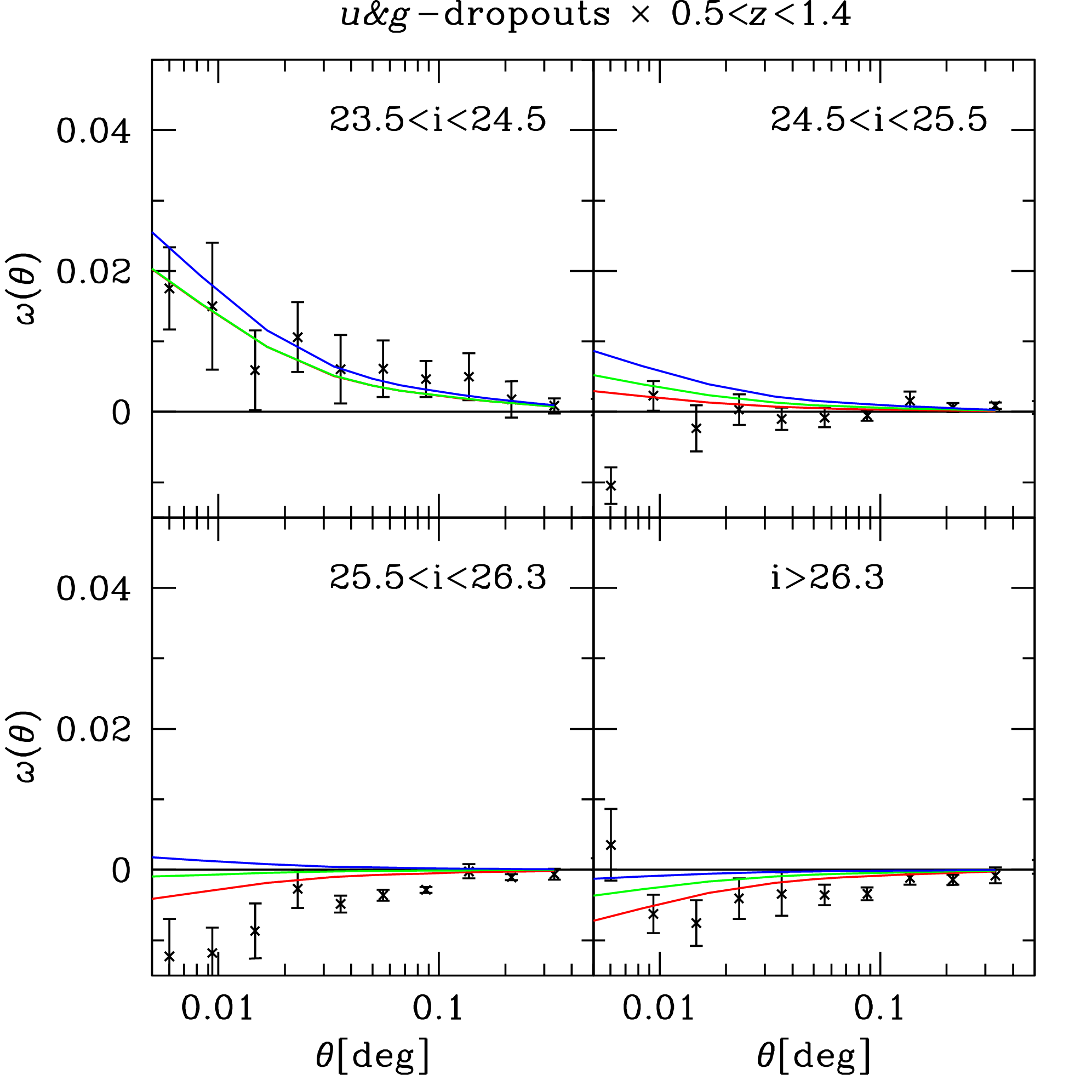}
\end{center}
\caption{Demonstration of magnification as a weak lensing measure, from \cite{hildebrandt2009}. Each panel shows the cross-correlation between foreground lenses and background galaxies identified with a dropout technique. Notice how in each case the brighter sources show a positive correlation, but the fainter ones anticorrelate, as expected from the shape of their luminosity function.}
\label{fig:mag2}
\end{figure}

 Magnification is in principle not possible to measure for an
individual source whose intrinsic brightness is unknown. (In strong
lensing the flux ratio of two images of the same source provides some
of this information, but in the weak lensing regime there is only one
image). However the brightness distribution of a population of sources
can be measured, as can its change due to lensing. Magnifying a piece
of sky containing a population of sources with a steep luminosity
function will boost the individual source fluxes, but will also dilute
the sky density by the same factor. For a sufficiently steep LF the
boost will win, and the net result will be an increase
in the number of sources above a given flux limit. For a flatter
luminosity function, on the other hand, the dilution effect dominates
and the number of sources will actually decrease as a result of the
magnification (Fig.~\ref{fig:mag1}).

This effect is difficult to apply to individual cluster lenses because
it can be affected by clustering of the background source
populations. But it has recently been demonstrated to work very nicely
on foreground galaxy lenses in the deep fields of the
Canada-France-Hawaii Legacy Survey \cite{hildebrandt2009}, using
$u$, $g$, and $r$ dropout galaxies as the sources: brighter dropouts
tend to show a positive correlation with the foreground sources,
whereas the fainter ones (which have a flatter luminosity function)
show an anticorrelation. Also qualitatively the strength of the signal
is as expected (Fig.~\ref{fig:mag2}).

A second effect is {\em flexion} \cite{baconetal2006}, the next-to-leading order distortion
after shear, due to the fact that the shear will be different in
different parts of a source. Flexion gives rise to the arc-like shapes
of lensed galaxies, for example. While this is a weaker effect than
shear, and it drops off faster with distance to the lens, it is a
valuable addition of information because it probes the inner regions around lenses
where there are not many background sources. Flexion also has the
advantage that, unlike shear, it distorts galaxies into `unnatural'
shapes, so that the shape noise is considerably lower than it is for
shear.

\section{The KiDS project}

After many years of waiting, we are finally looking forward to the VLT
Survey Telescope on Paranal being completed. It has been designed with
image quality over a wide field in mind, and even though it is late to arrive it will still provide a
unique capability in the Southern hemisphere. The hope (and design) is
that the telescope will deliver natural seeing images over a full
square degree field of view, well sampled with the small pixels of the
OmegaCAM camera that was built by a consortium of institutes in the
Netherlands (NOVA), Germany (G\"ottingen, Bonn, M\"unchen and ESO) and
Italy (Naples and Padua).

\begin{figure}
\begin{center}\includegraphics[width=\textwidth]{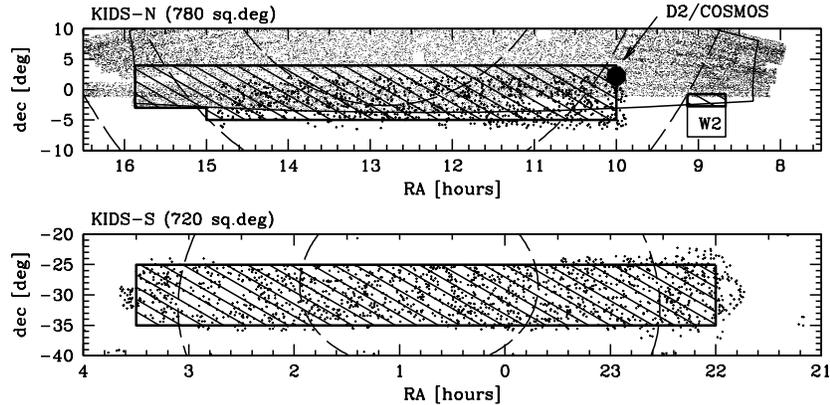}\end{center}
 \caption{Field layout of the KiDS survey. The hashed areas indicate the survey area to be imaged, 1500 square degrees in total. Small and large dots delineate the areas where massive spectroscopic surveys of the brighter galaxies have taken place with SDSS and 2dF respectively.}
\label{fig:kids}
\end{figure}

 The system allows an unprecedented weak lensing survey to be carried
out, and this is the aim of the largest survey, KiDS (the Kilo-Degree
Survey) that is planned for the VST as soon as it enters
operations (Fig.~\ref{fig:kids}). Together with a companion survey on the near-IR VISTA
survey telescope, the project will make images of 1500 square degrees
of sky, in nine photometric bands from $u$ to $K$ (Table~\ref{tab:kids}). The survey area includes many galaxies with known redshift, ensuring that the foreground mass distribution is mapped in detail and can be `weighed' with gravitational lensing. Because all
observations will be done in queue-scheduling, it will be possible to
optimize the use of the seeing distribution, so that the best seeing
dark time will be used for deep $r$ band exposures suitable for weak
lensing measurements, with brighter or
worse seeing time being devoted to the other photometric bands. As a
result the survey should be able to deliver high-fidelity weak
lensing {\em and} photometric redshift measurements over a large part
of the sky, with homogeneous data quality. In terms of the usual
tradeoff between area and depth, the survey sits between the wide and
shallower SDSS images (9000 square degrees) and the 170-square degree
CFHTLS. But the unique combination of image quality and 9-band photometry
should make it a very valuable resource. 

\begin{table}
\begin{center}
\caption{Main parameters of the KiDS+VIKING survey}
\label{tab:kids}
\begin{tabular}{crccc}
\hline\noalign{\smallskip}
band & exp.time & seeing ($''$) & 
5-$\sigma$ $2''$ limit & telescope\\
\noalign{\smallskip}\svhline\noalign{\smallskip}
$u$  &900 & 0.85-1.1  &  24.8  & VST \\
$g$  &900 & 0.7-0.85  &  25.4  & VST \\
$r$  &1800 & $<0.7$  &  25.2  & VST \\
$i$  &1080 & 0.7-1.1  &  24.2  & VST \\
$Z$  &500 &   $<1$   &  23.1  & VISTA \\
$Y$  &400 &    $<1$   &  22.4  & VISTA \\
$J$  &400 &    $<1$    &  22.2  & VISTA \\
$H$  &300 &   $<1$    &  21.6  & VISTA \\
$K$  &500 &    $<1$   &  21.3  & VISTA \\
\noalign{\smallskip}\svhline\noalign{\smallskip}
\end{tabular}
\end{center}
\end{table}

Science goals for the survey are broad, but the design 
centered on using weak lensing to measure the mass distribution in and around galaxies as a
function of environment, luminosity, and type, out to a radius of
several 100 kpc. This is an interesting regime because it is far
beyond the baryon-dominated regions of the galaxy and so can be
compared to robust predictions from structure formation theory and
simulations. A particularly interesting measurement will be the
average flattening of galaxy haloes from edge-on disk galaxies: most
alternatives to dark matter should make robust predictions for the
variation of the quadrupolar field with radius, and these can be
tested directly with such data.

The survey will also serve as a pathfinder for yet more ambitious lensing surveys for cosmology, including dark energy experiments from the ground and from space.

\section{Conclusions}

Galaxies provide one of the best environments for studying the dark matter. In the parlance of this conference, they are the mask that covers the mystery, but that also shows us where to look. 

Gravitational lensing is one of the most promising ways of studying the dark matter distribution, particularly at large radii where no classical dynamical tracers are available. New surveys such as KiDS will be able to map the dark matter in and between galaxy halos with great accuracy, which will enable new tests of structure formation as well as the cosmological model itself.

\begin{acknowledgement}
It is a pleasure to acknowledge the important role that Ken Freeman has played throughout my career as an astronomer, first as the author of a number of seminal papers, later as a source of occasional excellent advice, and now as a collaborator in the PN.S project, but foremost as a great person to be around. He and David Block organized a fantastic, memorable conference at Sossusvlei.

This paper benefited from the collaboration and hard work of my students and postdocs, in particular Jelte de Jong, Tim Schrabback and Hendrik Hildebrandt.
\end{acknowledgement}


\begin{thebibliography}{99.}

\bibitem{macho1993} C. Alcock, et al., 1993, Nature, \textbf{365}, 621.
\bibitem{eros1998} C. Alcock, et al., 1998, ApJ, \textbf{499}, L9.
\bibitem{baconetal2006} D.J. Bacon, D.M. Goldberg, B.T.P. Rowe, A.N. Taylor, 2006, MNRAS, \textbf{365}, 414.
\bibitem{binneymamon1982} J.J.Binney, G.A. Mamon, 1982, MNRAS, \textbf{200}, 361.
\bibitem{dejongetal2006} J.T.A. de Jong, et al., 2006, A\&A, \textbf{446}, 855.
\bibitem{fischer2000} P. Fischer, et al., 2000, AJ, \textbf{120}, 1198.
\bibitem{freeman1970} K.C. Freeman, 1970, ApJ, \textbf{160}, 811.
\bibitem{gavazzi2007} R. Gavazzi, et al., 2007, ApJ, \textbf{667}, 176.
\bibitem{hildebrandt2009} H. Hildebrandt, L. van Waerbeke, T. Erben, 2009, A\&A, \textbf{507}, 683.
\bibitem{hoekstra2004} H. Hoekstra, H.K.C. Yee, M.D. Gladders, 2004, ApJ, \textbf{606}, 67.
\bibitem{slacs3} L.V.E. Koopmans, T. Treu, A.S. Bolton, S. Burles, L.A. Moustakas, 2006, ApJ, \textbf{649}, 599.
\bibitem{mandelbaum2006a} R. Mandelbaum, U. Seljak, G. Kauffmann, C.A. Hirata, J. Brinkmann, 2006, MNRAS, \textbf{368}, 715.
\bibitem{mandelbaum2006b} R. Mandelbaum, C.A. Hirata, T. Broderick, U. Seljak, J. Brinkmann, 2006, MNRAS, \textbf{370}, 1008.
\bibitem{schrabbacketal2010} T. Schrabback, et al., 2010, A\&A in press. (arXiv:0911.0053)

\end{thebibliography}
\end{document}